%Paper: astro-ph/9206005
%From: PH520010@brownvm.brown.edu
%Date: Wed, 24 Jun 92 03:35:05 EDT

\input phyzzx
\hsize=6.5in
\hoffset=0.0in
\voffset=0.0in
\vsize=8.9in
\FRONTPAGE
\line{\hfill BROWN-HET-859}
\line{\hfill June 1992}
\vskip1.2truein
\titlestyle{Entropy of a Classical Stochastic Field}
\titlestyle{and Cosmological Perturbations}
\bigskip
\author{R. Brandenberger$\sp{1),\thinspace 3)}$, V. Mukhanov$\sp{2),}$
    \footnote{*}{on leave of absence from Institute for Nuclear Research,
Academy of Sciences, 117312 Moscow, Russia}
           and T. Prokopec$\sp{1),\thinspace 3)}$}
%\andauthor{T. Prokopec$\sp{1)}$}
\medskip
\centerline{1) \it Physics Department, Brown University}

\centerline{\it Providence, RI~02912, USA}
\medskip
\centerline{2) \it Institute of Theoretical Physics, ETH H\"{o}nggerberg,}

\centerline{\it CH -- 8093 Z\"{u}rich, Switzerland}
\medskip
\centerline{3)\it Institute for Theoretical Physics, University of California}

\centerline{\it Santa Barbara, CA 93106, USA}
\medskip
%\bigskip
\abstract
We propose a general definition of nonequilibrium entropy
of a classical stochastic field. As an example of particular interest in
cosmology we apply this definition to compute the entropy of density
perturbations in an inflationary Universe. On the scales of
structures in the Universe, the entropy of density perturbations
dominates over the statistical fluctuations of the entropy of cosmic
microwave photons, indicating the relevance of the entropy of
density fluctuations for structure formation.
\endpage
\chapter{Introduction}
\par
\REF\Huang{See {\it e.g.\/} K.~Huang, {\it Statistical Mechanics\/}
              (Wiley, New York, 1963); \nextline
               R.~Kubo, {\it Statistical Mechanics\/}
              (Elsevier, Amsterdam, 1965).}
The concept of entropy contains relevant information about a dynamical
system\rlap.\refmark{\Huang}\
In systems with a finite number of degrees of freedom
there is a natural way to define entropy, even if the system is out of
thermal equilibrium. We are interested in systems with infinitely
many degrees of freedom and which can be described by a stochastic
Gaussian field. In this letter, we propose a general definition
of nonequilibrium entropy of a stochastic Gaussian field.
\par
\REF\Bekenstein{J.~Bekenstein, {\it Phys. Rev.\/}~{\bf D\thinspace 7,}
                            2333\thinspace(1973).}
\REF\Hawking{S.~Hawking, {\it Comm. Math. Phys.\/}~{\bf 43}\thinspace,
               199\thinspace (1975).}
\REF\Penrose{R.~Penrose, in {\it `General Relativity\thinspace: An Einstein
                 Cetenary Survey'\thinspace,\/}~ed. by S.~Hawking and W.~Israel
                        (Cambridge University Press, Cambridge, 1979).}
\REF\Smolin{L.~Smolin, {\it Gen. Rel. Grav.\/}~{\bf 17},~417\thinspace
                       (1985); \nextline
B.~Hu and D.~Pavon, {\it Phys. Lett.\/}~{\bf 180\thinspace B}\thinspace,
                     329\thinspace (1986);\nextline
H.~Kandrup, {\it J. Math. Phys.\/}~{\bf 28}\thinspace, 1398
                     \thinspace(1987);\nextline
B.~Hu and H.~Kandrup, {\it Phys. Rev.\/}~{\bf D\thinspace35\thinspace,}
                      1776\thinspace (1987);\nextline
H.~Kandrup, {\it Phys. Rev.\/}~{\bf D\thinspace37}\thinspace,
                       3505\thinspace (1988);\nextline
E.~Calzetta and B.~Hu, {\it Phys. Rev.\/}~{\bf D\thinspace37\thinspace,}
                       2878\thinspace (1988).}
An issue of considerable interest is to develop a consistent
definition of entropy in general relativity and cosmology. There have
been some key results in this area. The observation that all information
about a particle crossing the Schwarzschild horizon is lost, led
Bekenstein\refmark{\Bekenstein}\
and Hawking\refmark{\Hawking}\ to their famous formula for the
entropy of a black hole.
Penrose\refmark{\Penrose}\
suggested that it may be possible
to realize the second law of thermodynamics in cosmology by assigning
an entropy to the gravitational field itself. He conjectured that the
plausible definition of entropy might be ``some integral'' of the Weyl
tensor squared, and that the Universe starts in a state of minimal
gravitational entropy. In this picture structure formation
and the second law of thermodynamics are reconciled, since
gravitational clustering leads to an increase in the Weyl tensor,
thus generating gravitational entropy. The connection between
information loss and entropy of the gravitational field
was explored in many papers\rlap.\refmark{\Smolin}
\par
In this letter we use our definition of nonequilibrium entropy of a
stochastic Gaussian field to propose a new approach to the problem of
gravitational entropy in cosmology. The formalism is based on
separating the entire system of gravitational plus matter fields
into background fields (chosen to have high space-time symmetries) and
linearized fluctuating fields; the latter are the stochastic fields we apply
our general definition of entropy to.
\par
\REF\Mukhanov{V.~Mukhanov, H.~Feldman and R.~Brandenberger,
{\it Theory of Cosmological Perturbations\thinspace,\/}~{\it Phys.
Rep.\thinspace,\/}
 in press (1992).}
It has been demonstrated that the dynamics of perturbations can be
reduced to the
dynamics of a single scalar field (which comprises in a self-consistent manner
both scalar and/or tensor gravitational field perturbations and matter field
fluctuations) in the classical space-time background (for a recent
review see Ref.~[\Mukhanov]\thinspace).
The evolution of the
background field is completely
specified; this means it carries {\it no} entropy.  On the other hand,
the fluctuating field carries significant entropy.  This statement needs
justification.
\par
\REF\Parker{L.~Parker, {\it Phys. Rev. Lett.\/}~{\bf21\thinspace,}
            562\thinspace (1968);\nextline
            L.~Parker, {\it Phys. Rev.\/}~{\bf183\thinspace,}
            1057\thinspace (1969);\nextline
            L.~Grishchuk, {\it Zh. Eksp. Teor. Fiz.\/}~{\bf 67\thinspace,}
            825\thinspace (1974).}
In order to obtain growth of entropy, it is necessary to propose some kind of
coarse graining in which some information is lost
during the evolution.  In this work, we consider a free scalar field in an
expanding space-time background, in which there is abundant production of
perturbations by parametric
amplification\rlap.\refmark{\Parker}\
We assume that
there is a mechanism which generates stochasticity in the
phases of the perturbations
produced during the evolution.  This mechanism is effective for the
field modes within the horizon, and generates entropy in the fluctuating
field.
\par
The letter is organized in the following way.  The next section is devoted to
the derivation of the formula for the entropy of a stochastic Gaussian scalar
field.  Using this formula, we then calculate (in Sec.~3.) the entropy of
cosmological density perturbations.
% structure formation.
In Sec.~4. we hint to some additional possible applications of the
formula in the context of cosmology.
\chapter{Entropy of Classical Field}
\par
We wish to consider the entropy associated with a classical stochastic
field.  For a given real scalar field
$\phi$~and its canonical momentum $\pi$\thinspace, there is a
probability distribution functional $P [\varphi, \pi]$  defined over
an infinite dimensional space spanned by functions $\{ \varphi, \pi
\}$.  The probabilistic definition of entropy gives
$$
S = -\int P [ \varphi, \pi ] \, \ln \, P [ \varphi, \pi ] \, {\cal D} \varphi
\, {\cal D} \pi \, , \eqno\eq
$$
where the probability functional $P [ \varphi, \pi]$ is normalized to unity.
We assume a Gaussian process,~{\it i.e.\/} that the knowledge of two-point
correlation functions suffices to completely specify the stochastic
properties of the
fields $\varphi$ and $\pi$.  (Higher order correlation functions can be then
given in terms of two-point correlations.)
\par
If the stochastic process is non-Gaussian, then the Gaussian approximation
may still be
a good one, provided corrections due to higher order correlations are small.
An additional requirement is that correlations are of finite range.  This is
fulfilled in the cosmological set up, because a natural cut-off for
correlation is the horizon scale.
\par
The Gaussian approximation breaks down when perturbations grow nonlinear, and
effects of the nonlinearities (originating in the full theory) become
significant.  In this case the corrections arising in the higher order
correlation functions become important and need to be incorporated in
the probability distribution.
\par
The definition of entropy in Eq. (1) can be applied to cosmological
perturbations in an expanding universe (see Sec.~3.), when the evolution of
perturbations can be well represented by a Hamiltonian of second order in
$\varphi$ and $\pi$.  This means that the Gaussian character of the
probability distribution is preserved in the course of evolution.
\par
We now sketch a derivation of the  expression for entropy in terms of
correlation functions.
Assume that at some time $t$ the probability functional $P [ \varphi,
\pi]$ has a general Gaussian form
$$
\eqalign{
P [\varphi, \pi] & = {1\over {\cal N}} \exp \Bigl[ - \int {1\over 2}
\bigl\{ \varphi
(\vec x) A (\vec x, \vec y) \varphi (\vec y)   \cr
&   + \pi (\vec x) B (\vec x, \vec y)
\pi (\vec y) + 2 \varphi (\vec x) C (\vec x, \vec y) \pi (\vec x, \vec y)
\bigr\} d^3 x d^3 y \Bigr]\, , } \eqno\eq
$$
where $\cal N$ is a normalization constant, and $A, \, B, \,$ and $C$ are
related
to the two-point correlation functions in a way yet to be determined. In a
homogeneous space-time background, $A, \, B,$ and $C$ are
functions of $(\vec x - \vec y)$ only.
\par
By a clever substitution, it is possible to bring Eq. (2.2) into a
diagonal form,
in which $\varphi$ and $\pi$ are replaced by new normal coordinates. It
is then quite straightforward to evaluate the normalization  factor
$\cal N$ of Eq.~(2.2)
$$
   {\cal N} = \sqrt{\det {\cal D} (\vec x - \vec y)} \, , \eqno\eq
$$
where ${\cal D}$ can be expressed in terms of correlation functions
$$
{\cal D} (\vec x - \vec y) = \int d^3 z \bigl[ \langle\varphi (\vec x) \varphi
(\vec z) \rangle \,\langle \pi (\vec z) \pi (\vec y)\rangle -
\langle \varphi (\vec x) \pi (\vec z) \rangle
\langle\pi (\vec z) \varphi (\vec y)\rangle \bigr] \, . \eqno\eq
$$
Using Eqs.~(2.2) through~(2.4), the expression~(2.1) for the entropy gives
$$
S = Tr \, \delta (\vec x - \vec  y) + \ln \, {\cal N} \, . \eqno\eq
$$
The first term is an irrelevant constant.
The relevant contribution comes from the second term and can be rewritten as
$$
S = {1\over 2}\ln {\det \, {\cal D} (\vec x - \vec y) } \, . \eqno\eq
$$
The above equation is the main result of this section.  Eqs.~(2.4) and~(2.6)
can
be used to obtain the entropy of any stochastic classical scalar field whose
probability distribution can be well approximated by the Gaussian
probability distribution~(2.2).
\par
\REF\Brandenberger{R. H. Brandenberger, V. Mukhanov and T. Prokopec,
{\it The Entropy of Gravitational Perturbations\/}, Brown University
preprint, BROWN-HET-849\thinspace(1992).}
In order to calculate the determinant of ${\cal D}(\vec x, \vec y)
\thinspace,$
one needs to solve the eigenvalue problem associated with $\cal D$.
 Under quite general conditions (assuming ${\cal D}$ is
of finite support) and using the $\zeta$-function regularization scheme, it is
possible to show\refmark{\Brandenberger}~
that this determinant can be expressed in terms of the
spectral density ${\cal D}_{\vec k}$\thinspace, which is given by the
Fourier transform of ${\cal D} (\vec x - \vec y)$
$$
{\cal D}_{\vec k} \equiv \int d^3 z\,
e^{-i \vec k \cdot \vec z} {\cal D} (\vec z) =
\langle\vert \varphi_{\vec k}\vert^2 \rangle\,\langle\vert
 \pi_{\vec k}\vert^2 \rangle - \vert\langle \varphi_{\vec k}
\pi_{-\vec k} \rangle\vert^2 \,                          \eqno\eq
$$
and it is positive definite.
The entropy then reads
$$
S = V \int \, {d^3 k\over{(2 \pi)^3}} \, {1\over 2}\ln {{\cal D}_{\vec k}}
 \, . \eqno\eq
$$
\par
The procedure to calculate the entropy of a classical Gaussian field
is now very simple.  Given two point correlation functions, one calculates
${\cal D} (\vec x - \vec y)$ (Eq.~(2.4)), Fourier transforms it
(Eq.~(2.7)) and obtains the entropy according to
Eq.~(2.8).  We now apply this prescription to an example which is of
interest in  cosmology.
\par
\chapter{Entropy of Cosmological Perturbations}
\par
In this section we apply the method developed above to  calculate
the entropy of
cosmological density perturbations.  This is an example of relevance in
cosmology, because it is likely that the scalar density perturbations seed
structures in the universe.  We find that the entropy of scalar density
perturbations on large scales in the universe is significant when compared
to the statistical fluctuations of the entropy of cosmic microwave
photons on the same scales.
\par
\REF\BrandenbergerB{R.~Brandenberger, H.~Feldman and V.~Mukhanov,
      {\it Gauge Invariant Cosmological Perturbations\thinspace,\/} Brown
       preprint BROWN-HET-841\thinspace(1992)\thinspace,
       to be published in the
       {\it Proceedings of ICGC-91\/} (Wiley Eastern Ltd.\thinspace,
        New Delhi, 1992\thinspace).}
Before we present any calculations, we give a short summary of the theory of
density perturbations.  Density perturbations are scalar type metric
perturbations which couple to energy density and pressure.  For matter which is
in the form of a scalar field, or an ideal gas, it turns out that
density perturbations can be described in a self-consistent manner in terms of
a single gauge invariant scalar field $\varphi$, which is a
linear combination of scalar field matter fluctuations (ideal gas density
fluctuations) and longitudinal metric fluctuations and
whose dynamics is given by a quadratic action.  (For a comprehensive account
of the gauge invariant formalism of linear cosmological perturbation in
Friedmann-Robertson-Walker backgrounds see Ref.~[\Mukhanov]\thinspace,
for pedagogical introduction see Ref.~[\BrandenbergerB\thinspace]\thinspace.)
\par
  Hence, assuming that the
gauge invariant field $\varphi$ is a stochastic Gaussian field, we can
calculate the
entropy associated with $\varphi$ using Eqs. (2.4) and (2.6), or equivalently
Eqs. (2.7) and (2.8). In order to accomplish this,
we need to know the two-point correlation functions
$\langle \varphi (\vec x, t) \varphi (\vec y, t) \rangle, \,
\langle \pi (\vec x, t) \pi (\vec y, t) \rangle$, and
$\langle \varphi (\vec x, t) \pi (\vec y, t)\rangle$, where $\pi (\vec x, t)
= {\partial\over{\partial t}} \varphi (\vec x, t)$.
\par
As an example of a model in which the correlation functions exhibit nontrivial
behavior, we consider an expanding universe with initial quantum
fluctuations which evolve into classical ones as a result of evolution.
In this case, particle
pairs are produced via parametric amplification, i.e., via coupling of matter
fields to the nontrivial space-time background.  Because of the generation of
perturbations, the correlation functions become time dependent.  Here we
consider
the inflationary universe scenario in which there is abundant
production of inhomogeneities.
\par
The Hamiltonian governing the evolution of the single scalar field
$\varphi$ and momentum $\pi$ is quadratic in $\varphi$ and $\pi$, so that it
is convenient to represent the evolution operator $\hat U(t)$
in a form in which the effects of free evolution $(\cal R)$ and interaction
with the background $(S)$ are separated. $\hat U(t)$ is the product of the
rotation operator $\cal R$ and the two-mode squeeze operator $S$
$$
\hat U (t) = {\cal R} ( \{ \theta_{\vec k} \}) \>
               S( \{ r_{\vec k} , \varphi_{\vec k} \} ) \eqno\eq
$$
where
$$
\eqalign{ {\cal R} ( \{ \theta_{\vec k} \} ) & = \prod_{\vec k, \, k_x > 0}
{\cal R} ( \theta_{\vec k} ) \cr
S ( \{ r_{\vec k} , \varphi_{\vec k} \} ) & = \prod_{\vec k, \, k_x > 0}
S (r_{\vec k} , \varphi_{\vec k} ) \, . } \eqno\eq
$$
\par
\REF\Caves{C.~M.~Caves and B.~L.~Schumaker,
{\it Phys. Rev.\/}~{\bf A\thinspace31}\thinspace, 3068 and 3093
\thinspace (1985).}
The product $ \prod_{\vec k, \, k_x > 0} $ is over half of
the possible values of momenta $\vec k$ (for definiteness, say $k_x > 0$).
The rotation angles $\theta_{\vec k} = \int^{t} \omega_{\bar k}(t') dt'$
are given in terms of the frequency $\omega_{\vec k}$ of mode
$\vec k$; $r_{\vec k}$, $\varphi_{\vec k}$ are the squeeze factor and phase,
respectively, and can be expressed in terms of parameters of
the Hamiltonian (see Ref.~[\Caves]\thinspace).
The two-mode squeeze operator $S (r_{\vec k}, \,
\varphi_{\vec k})$ acts on the vacuum
$\ket{0_{\rm in}}\thinspace,$ creating pairs of particles with momenta
$\vec k$ and $-\vec k$, so that the total momentum of the pair is zero;
hence $S(r_{\vec k}, \, \varphi_{\vec k})$ is a momentum conserving operator.
 The operator $S$ mathematically describes the process of parametric
amplification.
\par
\REF\Prokopec{T.~Prokopec, {\it Entropy of the Squeezed Vacuum\thinspace,\/}
Brown preprint, BROWN-HET-861\nextline (1992).}
Now we can express the two point correlation functions of quantum operators
$\hat \varphi$ and $\hat \pi$ in terms of the parameters of the squeezed
state which is obtained as a result of the evolution of the initial
vacuum state $\ket{0_{\rm in}}$ of cosmological perturbations in an
expanding Universe. Simple, but rather lengthly calculation
gives\refmark{\Prokopec}
$$
\eqalign{ \bra{0_{\rm in}} \varphi (\vec x , t) \varphi (\vec y, t)
\ket{0_{\rm in}}  & = \int
{d^3 k\over{(2 \pi)^3}} e^{-i \vec k \cdot (\vec x - \vec y)} \,
{1\over{2 \omega_{\vec k} (t) }} \cr
& \Bigl( (2 \sinh^2 r_{\vec k} + 1) - \sinh 2 r_{\vec k} \cdot \cos 2 \bigl(
\int \omega_{\vec k} dt - \varphi_{\vec k} \bigr) \Bigr) \, . }
\eqno\eq
$$
where the frequencies $\omega_{\vec k}$ and the squeeze factor $r_{\vec k}$
depend on time because of the nontrivial evolution of the background.
Similar expressions are obtained for
$\bra{0_{\rm in}}\pi (\vec x, \, t) \, \pi \,
(\vec y, \, t)\ket{0_{\rm in}}$ and
$\bra{0_{\rm in}} \varphi (\vec x, \, t) \,
\pi (\vec y, \, t)\ket{0_{\rm in}}$.
\par
Now we argue that the contribution due to the second (oscillating)
term of Eq.~(3.3)
can be neglected in the classical limit, when the number of
produced particles per mode is large. This can be justified if we are
considering the transition from quantum fluctuations to classical ones
and performing some coarse graining. The stochasticity which is
induced in the phases by coarse graining causes strong cancellation
in the oscillating terms.
\par
As a result we obtain for the spectral density of the operator
${\cal D}(\vec x - \vec y)$
$$
{\cal D}_{\vec k} = \sinh^2 \, r_{\vec k} \, (1 + \sinh^2 r_{\vec k})
                      \, .                                       \eqno\eq
$$
The entropy density per mode is then (Eq. (2.6))
$$
s_{\vec k} ={{S_{\vec k}}\over V}
= {1\over 2}\ln \, {\cal D}_{\vec k} \eqno\eq
$$
and ${\cal D}_{\vec k}\approx n\sp{2}_{\vec k}=\sinh\sp{2}\, r_{\vec k}\, $
(for $n_{\vec k}\gg 1\, $), where $n_{\vec k}$
is the average
number of particles in  $\vec k$-mode,
%($n_{\vec k}\approx \sinh\sp{2}\, r_{\vec k}$ for $n_{\vec k}\gg 1$),
whenever the notion of particles can be defined.
Note that Eq.~(3.5) can be applied even if the notion of particle
is not well defined. In particular this is the case for
inhomogeneities in a matter dominated Universe.\refmark{\Mukhanov}~
It is useful to define the entropy density
$s_{\lambda}$
per logarithmic wavelength $\lambda\sim{1/k}$ interval
$$
  s_{\lambda}\sim k\sp{3}\, s_{\vec k}
               \approx {1\over{\lambda\sp{3}}}\, \ln{{\cal D}_{\vec k}}\, .
                          \eqno\eq
$$
\par
\REF\Linde{A.~Linde, {\it Phys. Lett.\/}~{\bf 129\thinspace B}\thinspace,
177\thinspace (1983)\thinspace.}
To demonstrate how the technique developed above works, we now apply
Eq.~(3.5) to estimate the entropy of cosmological perturbations
produced during the inflationary stage of a model of chaotic
inflation\thinspace\rlap.\refmark{\Linde}\
The simplest potential for the inflaton field
$\phi_{\rm \, I}$ is $V(\phi_{\rm \,I})=(1/2)m\sp{2}\phi\sp{2}_{\rm \, I}$
\thinspace,
where $m$ is the mass of the inflaton, typically of the order
$10\sp{13}\hbox{\rm GeV}$. Considering perturbations on scales which enter
the horizon late in the radiation era,
we obtain the following result for the entropy density
of perturbations on a typical scale $\lambda_{\rm ph}\sim a/k$
(see Ref.~[\Brandenberger\thinspace]\thinspace)
$$
  s_{\lambda_{\rm ph}} \sim {1\over{\lambda\sp{3}_{\rm ph}}}\,
        \ln \, \Bigl[\, (m\, l) \,
\ln{{\lambda_{\rm ph}}\over{\lambda_{\gamma}}}\,
\bigl({{\lambda_{ph}} \over {t}} \bigr) \bigl( {{\lambda_{ph}} \over {l}}
\bigr) \Bigr]\, , \eqno\eq
$$
where $l$ is the Planck length,
$t$ is the cosmological time and $\lambda_{\gamma}$ is the
typical wavelength of the cosmic microwave background radiation.
This formula is applicable for
perturbations which satisfy the condition
$t\gg\lambda_{\rm ph}\, \gg \lambda_{\gamma}$\thinspace. The contribution
to the total entropy of galactic scale perturbations
($\sim 1 - 100\, \hbox{\rm Mpc}$) per corresponding galactic volume is
$S_{\rm gal}\sim \lambda\sp{3}_{\rm gal}\cdot s_{\rm gal}\approx
\hbox{200 - 220}\, .$
% For cosmic microwave photons,
%$S_{\rm gal}\approx \ln\, \lambda_{\rm gal}\, T_{\gamma}\sim 50\, ,$
%where $T_{\gamma}$ is the temperature of the microwave background radiation.
The entropy of gravitational radiation can be estimated in the same
manner (see Ref.~[\Brandenberger\thinspace]\thinspace)\thinspace.
\par
Statistical fluctuations of cosmic microwave photons are another potential
source of inhomogeneities. However, the entropy
density of these fluctuations scales
as $\lambda^{-3/2}$ compared to the logarithmic dependence we found for
the entropy density of density perturbations. Hence, on scales of galaxies,
the entropy of gravitational
perturbations dominates over the statistical fluctuations
of the entropy of cosmic microwave photons.
The total entropy of cosmological fluctuations is, however, suppressed
by a factor $(H/m_{\rm pl})\sp{3/2}$ (where $H$ is the Hubble expansion
rate at the end of inflation) compared to that of the cosmic
microwave background. The dominance of the entropy of gravitational
perturbations on large scales is a sign of the relevance of this
entropy for structure formation. A further
application\refmark{\Brandenberger}\ of this
entropy is in the context of a collapsing Universe.
\par
It is worth noting that because of the weak (logarithmic) dependence of
entropy per mode on the energy scale $m$ of the model, our conclusion
remains valid for a wide class of inflationary models.
\chapter{Discussion}
\par
We derived a new formula for the entropy of a stochastic Gaussian scalar
field in terms of two-point correlation functions.
We then applied this result to the cosmologically relevant example
of density perturbations, using the formalism of linearized gauge invariant
scalar perturbations about a homogeneous classical space-time and matter
background.  We found that the entropy of the system (scalar gravitational
metric
perturbations plus matter density fluctuations) grows as a logarithm of
the number
of particles created as the universe expands.  On the scales of large
structures in the universe, the entropy of density perturbations in an
inflationary Universe dominates over the entropy of statistical fluctuations
of the cosmic microwave photons.
\par
The formalism for calculation of entropy which we developed
in this letter can be applied to any
(cosmological) problem, which can be reduced to the evolution of
a classical stochastic Gaussian  field.
\par
The treatment of the full (nonlinear) gravitational field is still an open
problem, and the corresponding formula for entropy is yet to be constructed.
It would be very instructive to demonstrate that the entropy of the
gravitational field continues to grow, even when the perturbations
become nonlinear.
\par
\ack
For interesting discussions we are grateful to Andy Albrecht,
Leonid Grishchuk, Jim Hartle,
Tony Houghton, Bei-Lok Hu and Henry Kandrup. Two of us (R.\thinspace B.
and T.\thinspace P.\thinspace) thank the ITP of the University of California
in Santa Barbara for hospitality during the completion of this work.
V.\thinspace M. thanks the Swiss National Science Foundation for
financial support. At Brown, this work was supported by DOE grant
\hbox{DE-AC02-76ER03130} Task A and by an Alfred P. Sloan Foundation
Fellowship to R.\thinspace B. At the ITP in Santa Barbara
financial support from
NSF grant \hbox{PHY89-04035} is acknowledged.
\par
\endpage
\refout
\end